# Is "Compressed Sensing" compressive? Can it beat the Nyquist Sampling Approach?

Leonid P. Yaroslavsky,

*Dept. of Physical Electronics, School of Electrical Engineering, Tel Aviv University,
Tel Aviv 699789, Israel*
*Corresponding author: yaro@eng.tau.ac.il

Data compression capability of "Compressed sensing (sampling)" in signal discretization is numerically evaluated and found to be far from the theoretical upper bound defined by signal sparsity. It is shown that, for the cases when ordinary sampling with subsequent data compression is prohibitive, there is at least one more efficient, in terms of data compression capability, and more simple and intuitive alternative to Compressed sensing: random sparse sampling and restoration of image band-limited approximations based on energy compaction capability of transforms. It is also shown that assertions that "Compressed sensing" can beat the Nyquist sampling approach are rooted in misinterpretation of the sampling theory.

This letter is motivated by recent OPN publications ([1], [2]) that advertise wide use in optical sensing of "Compressed sensing" (CS), a new method of image digital image formation that has obtained considerable attention after publications ([3] - [7]). This attention is driven by such assertions in numerous publications as "CS theory asserts that one can recover certain signals and images from far fewer samples or measurements than traditional methods use" ([6]), "compressed sensing theory suggests that one can recover a scene at a higher resolution than is dictated by the pitch of the focal plane array" ([8]), "Beating Nyquist with Light" ([1], [9]). For those who are familiar with sampling theory and know that the Nyquist rate can't be beaten these assertions sound questionable. Is that true that "Compressed sensing guarantees on reconstruction quality even when sampling is very far from Nyquist sampling", and, if yes, how much one can win in terms of reducing sampling rate?

In order to answer these questions, consider first what sampling theory tells and how ordinary sampling works. In their claims, publications on Compressed Sensing refer to the Kotelnikov's - Shannon's sampling theorem in its classical formulation: signals that has no frequencies outside the interval $[-B/2, B/2]$ can be precisely restored from their samples taken with a sampling interval $1/B$, i.e. with sampling rate $B$ called the Nyquist rate.

Kotelnikov-Shannon's sampling theorem is the fundamental theorem of the sampling theory, but it is not the whole theory. First of all, the most general formulation of the sampling theorem is the following: if total area occupied by signal spectrum in spectral domain is $B$, the signal can be precisely restored from sampled data collected with the rate $B$ samples per unit of signal area. Such signals are called band-limited. Second, no band-limited signals exist in reality, and the sampling theory is the theory of signal band-limited approximation optimal in terms of mean squared approximation error. Therefore, sampling interval, or, correspondingly, sampling rate, are not signal attributes, but rather a matter of convention on which reconstruction accuracy is acceptable.

For instance, sampling intervals for image sampling are conventionally chosen on the base of a decision on how many pixels are sufficient for sampling smallest objects or object borders to secure their resolving, localization or recognition. The sampling interval found in this way determines the effective image base band, which specifies, for instance, the bandwidth of signal analog amplifiers or transmission systems. Since the area occupied in images by the smallest objects and object borders is usually a very small fraction of the total image area, such uniform sampling produces very many redundant samples and Fourier, DCT, wavelet and some other transform power spectra of images quite rapidly decay on high frequencies. This property of signal/image spectra is what in CS community is called "signal sparsity". The capability of transforms to compress image energy in few spectral coefficients is called their energy compaction capability. It is used in transform signal coding for compressing signal discrete representation obtained with ordinary sampling, i.e. for substantial reduction of the quantity of data required for signal reproducing with a given acceptable quality. In transform signal coding, signals are replaced by their "band-limited", or "sparse" approximations, i.e. by their copies that contain only a limited number of non-zero transform components, fewer than the number of signal samples.

Compressed sensing also assumes signal "sparse" approximation and suggests another approach to signal discretization that, admittedly, avoids the need of subsequent compression. According to this approach, one should specify the total number $N$ of signal samples required for its discrete representation and the number $M < N$ of certain measurements to be done in order to obtain signal discrete representation with $N$ samples by means of signal $L1$ norm minimization in a selected transform domain. The ratio $CSCF = N/M$ defines the degree of the compression achieved, the Compressed Sensing Compression Factor.

Data compression capability of Compressed sensing can be evaluated with respect to signal sparsity, i.e. to the ratio $K/N$ of the number $K$ of non-zero transform coefficients of the signal "sparse" approximation to the total number of signal samples $N$.

The ultimate upper bound of the compression that can be achieved by signal "sparse" approximation can be evaluated using the Discrete Sampling Theorem ([13]), according to which if only $K$ out of $N$ discrete signal spectral components are non-zero, all signal $N$ samples can be precisely restored from its $K$ samples provided that indices of the non-zero spectral components are known and certain restrictions regarding positions of $K$ signal samples are met. The restrictions are defined by the signal transform. For such transforms as DFT and DCT, positions of known signal samples can be arbitrary ([13]). For other transforms, such as, for instance, wavelet transforms, there are certain limitations in this respect.

Because, according to the discrete sampling theorem, $K$ is the minimal number samples sufficient for reconstruction of signal $K$-sparse approximation, the ratio

$$CUB = 1/SPRS , \qquad (1)$$

inverse to the signal sparsity $SPRS = K/N$ is obviously the ultimate theoretical upper bound of signal compression that can be achieved by signal sparse approximation. This relationship is plotted in Figure 1 (solid line). It is sufficiently well fit by empirical data on compression factor of JPG high quality coding (solid dots) obtained, using Matlab encoding tool, for a set of test images listed in Table 1 (three left columns). Effective image sparsity for these images was evaluated, for each particular image, for RMS of sparse approximation error equal to its RMS JPG coding error.

The upper bound of compression $N/M$ of signals with sparsity $K/N$ achievable by "Compressed sensing" can be from the relationship

$$\frac{M}{N} > 2\frac{K}{N}\log(\frac{N}{M})(1 + o(1)) \qquad (2)$$

provided in Ref. [7]. For the evaluation of the multiplier $(1 + o(1))$ in Eq. 2 we'll make use of the obvious fact that for $K = N$ there must be $M = N$. Then the relationship (2) between signal sparsity $SSP = K/N$ and compressed sensing compression factor $CSCF = N/M$ can be written as

$$SSP < \frac{1}{2CSCF \log(CSCF) + 1} \qquad (3)$$

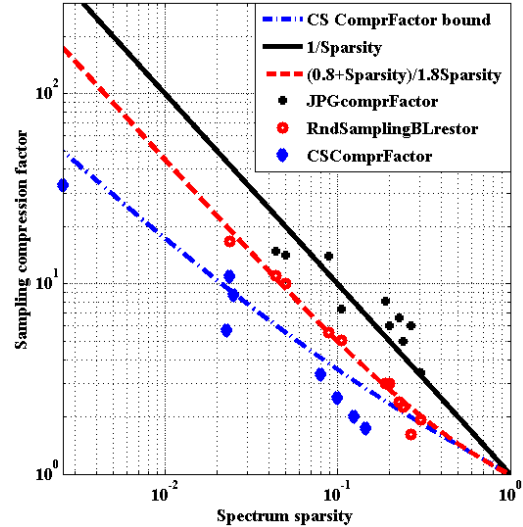

Figure 1. Sampling compression factor versus spectrum sparsity: theoretical upper bound (Eq. 1, solid), theoretical upper bound for Compressive sensing (Eq. 2, dash-dot) and estimate for random sampling and band-limited reconstruction (dash)

Table 1. Experimental data on image compression achieved by JPG coding and using RSBL approximation methods

| Test image | JPG compr factor | JPG RMS error | Spectrum Sparsity. | RSBR Compr. factor | RMS error |
|---|---|---|---|---|---|
| Mamm | 14.8 | 1.48 | 0.044 | 11 | 1.58 |
| Ango | 14.1 | 1.25 | 0.05 | 10.5 | 1.36 |
| Test4CS | 13.9 | 1.62 | 0.89 | 5.55 | 2.15 |
| Moon | 7.35 | 2.5 | 0.105 | 5 | 2.55 |
| Lena512 | 8.1 | 3.9 | 0.19 | 3 | 3.3 |
| Aerial photo | 6 | 4.5 | 0.2 | 3 | 4.76 |
| Man | 6.65 | 4 | 0.227 | 2.38 | 4.12 |
| ManSparse[6] | - | - | 0.024 | 16.7 | 5.6 |
| Multi-photon | 4.98 | 4.89 | 0.238 | 2.26 | 4.73 |
| Barbara | 6.04 | | 0.265 | 1.61 | 4.15 |
| Westconcord | 3.42 | 7.21 | 0.301 | 1.92 | 7.48 |

This relationship for the upper bound of compression achievable for CS is plotted in Figure 1 by the dash-dot line. Additionally, plotted are numerical experimental data on actual CS compression factors for particular signals and images found in the literature (diamonds). The corresponding sources and data are listed in
Table 2Table 2.

Table 2. Experimental data on signal compression achieved by using "Compressed sensing" methods

| Source | Spectrum sparsity | Compression factor |
|---|---|---|
| [6], Figure 3 | 64/512=0.125 | 512/256=2[1) |
| [6],Fig. 1 | 25000/(1024*1024)= 0.0238 | (1024*1024)/92000=10.92[1) |
| [1], p. 48 | 0.015*0.03=0.00045 | 1/0.03=33[1) |
| [10] | 25000/(1024*1024)= 0.0238 | (1024*1024)/(430*430)= 5.67 [2) |
| [10] | 25000/(1024*1024)= 0.0238 | (1024*1024)/(350*350)= 8.56 [3) |
| [11] | 6500/(256*256)=0.0992 | (256*256)/26000=2.52[1) |
| [12], p.1 | 128/1600=0.08 | 1600/481=3.33[1) |
| [12], p.2 | 75/512=0.146 | 512/301=1.73[1) |
| [1) No data on signal sparse approximation accuracy are available | | |
| [2)RMS error of image sparse approximation 10.9 gray levels | | |
| [3) RMS error of sparse image restoration 11.33 gray levels | | |

As one can see, Compressed sensing requires a substantially redundant number of data with respect to the theoretical bound defined by Eq. 1. The degree of the redundancy is plotted in Figure 2 as a ratio $CUB/CSCF$ (dash-dot line).

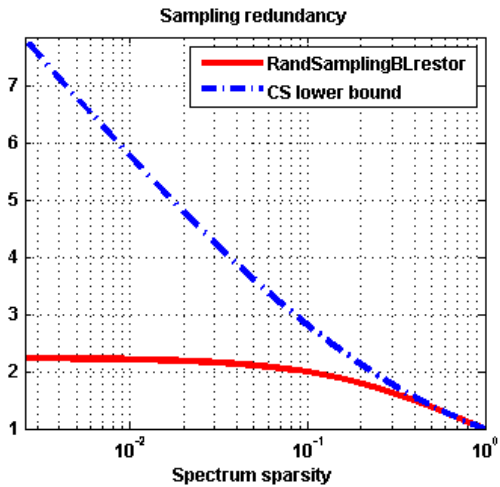

Figure 2. Redundancy of Compressed sensing and of Random sampling and band-limited approximation vs signal/image sparsity

This sampling redundancy of Compressed sensing is the price one should pay for the uncertainty regarding indices of signal non-zero spectral components.

Can one reach higher compression efficiency when sampling images than that of Compressed sensing? The answer is yes.

Consider one possible alternative. Compressed sensing approach to sampling assumes the belief that signals can be approximated by their "sparse", or band-limited copies. This belief is based on energy compaction capability of transforms. For overwhelming number of real images appropriate transforms, such as DCT, compact image energy into lower frequency part of spectral components. One can, therefore, in addition to specifying the number $N$ of desired images samples and the number $M$ of samples to be taken, which is anyway required by the CS approach, make a simple natural assumption that image spectral components important for image reconstruction are concentrated within, say, circular shapes that encompass $M$ image lowest frequency components. With this assumption, one can apply the discrete sampling theorem and reconstruct image band-limited approximation from a set of $M$ samples taken, in case of sparsity of DCT or DFT spectra, in randomly chosen positions ([13]).

Technically, this method can be implemented, for instance, using an iterative Papoulis-Gershberg type algorithm, which, at each iteration, sets to zero the selected spectral components and then restores, in image domain, pixels that were acquired at sampling.

This option, which we will call "Random Sampling and Band Limited Reconstruction" (RSBLR) is illustrated in Figure 3 on an example of a test image from a set of 11 test images listed in Table 1. Also listed in the table are numerical data obtained for all test images: (i) image spectrum sparsity on the level of RMS reconstruction error defined by the corresponding RMS JPG reconstruction error (fourth column), (ii) RSBLR compression factor (ratio of the image size, in pixels, to the number of random samples taken, fifth column) and (iii) RMS restoration error (in units of gray levels).

Obtained in these experiments values of compression factors for these images are plotted in Figure 1 (bold circles) along with the curve *(0.8+Sparsity)/(1.8\*Sparsity)* that fits them sufficiently well. Solid line in Figure 2 represents an estimate of the sampling redundancy of RSBLR obtained as a ratio of the theoretical upper bound of the possible compression factor for sparse approximation (Eq. 1) to the fitting curve *(0.8+Sparsity)/(1.8\*Sparsity)*. As one can see from Figure 2, Random Sampling and Band Limited Reconstruction is superior to Compressed sensing in its sampling compression efficiency in practically entire range of values of possible image sparsity.

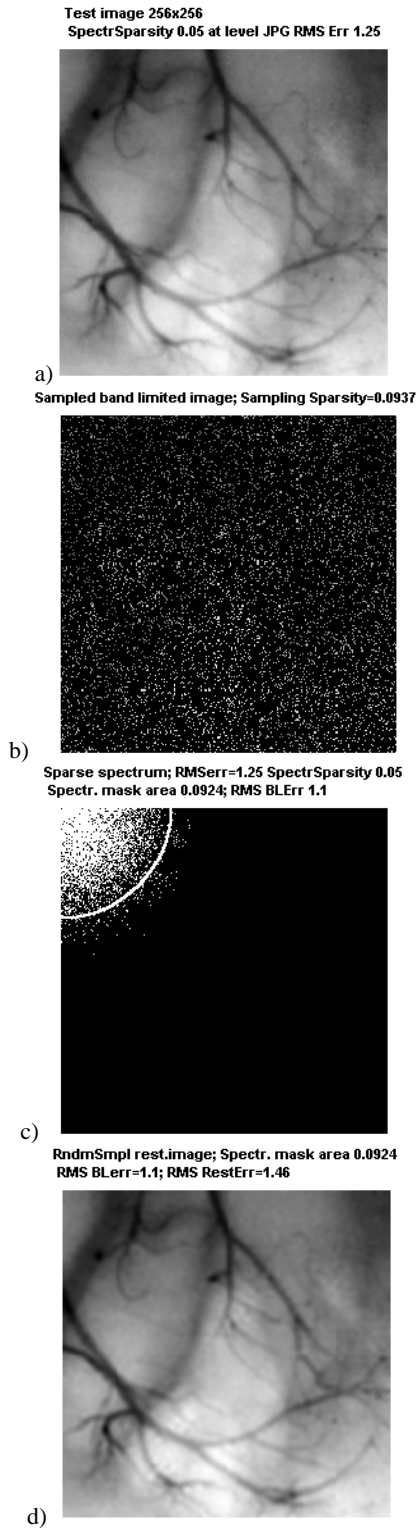

Figure 3. (a) - Test image of 256x256 pixels, (b)- its 6.141 samples taken in randomly selected positions, (c) - a map (white dots) of its 3277 most intensive DCT components, which reconstruct the image with the same RMS error of 1.25 gray levels as that of its JPG reconstruction error, and the border (white line) of the low pass filter that encompasses 6029 DCT spectral components; (d) - band-limited reconstruction with RMS error of 1.46 gray levels from the sparse random samples

Note that the RSBLR approach is straightforwardly applicable to image reconstruction from projections and to other digital imaging tasks as well.

Consider now assertions that "Compressed sensing" can "beat the Nyquist sampling approach". Compressed sensing technique can, in principle, restore signals with few spectral components within the base-band defined by the component of the highest frequency from their samples taken with a rate lower than twice this frequency. This, however, certainly does not mean at all that it beats the Nyquist sampling because twice the component highest frequency is not the Nyquist rate for such signals.

According to the sampling theory, sampling signals that contain few spectral components with sampling rate defined by the highest frequency component is too wasteful. For sampling such signals, the signal base-band should rather be split into sub-bands of the width of those sparse spectral components and sampling should be carried out of the sub-bands that have "non-zero" energy, i.e. energy above the measurement noise level. The effective sampling rate will then be defined only by the total area occupied by signal spectral components. Such optimal sampling requires signal sinusoidal modulation-demodulation in order to shift signal high frequency sub-bands to low frequency band before sampling and then to shift them back for signal reconstruction.

Compressed sensing replaces signal sinusoidal modulation-demodulation by signal blind modulation-demodulation using pseudo-random masks, but pays quite high price of substantial redundancy in the required number of samples. For instance, in the experiment of sampling and reconstruction of a high frequency sinusoidal signal presented in Ref. [1], the required redundancy (the ratio **M/K**, in denotations of the paper) is 1/0.015, i.e. about 67 times. Note that no analysis concerning the accuracy of signal high frequency sinusoidal components restoration and possible aliasing artifacts is provided in that publication, as well as in others similar.

To conclude, above presented theoretical estimates and experimental data show that assertions wide spread in the literature that CS methods enable large reduction in the sampling costs for sensing signals and surpass the traditional limits of sampling theory are quite over exaggerated and are rooted in misinterpretation of the sampling theory. It is shown also that there is at least one substantially less redundant and more simple and intuitive alternative to Compressed sensing for the cases when image regular sampling with subsequent data compression is prohibitive, that of random sparse sampling and restoration of image band-limited approximations based on a priori knowledge regarding the energy compaction capability of the transform.